# High Dynamic Range Pixel Array Detector for Scanning Transmission Electron Microscopy

Mark W. Tate[1], Prafull Purohit[1], Darol Chamberlain[2], Kayla X. Nguyen[3], Robert M. Hovden[3], Celesta S. Chang[4], Pratiti Deb[4], Emrah Turgut[3], John T. Heron[4,5], Darrell G. Schlom[5,6], Daniel C. Ralph[1,4,6], Gregory D. Fuchs[3,6], Katherine S. Shanks[1], Hugh T. Philipp[1], David A. Muller[3,6,*], Sol M. Gruner[1,2,5,6]

[1]Laboratory of Atomic and Solid State Physics, Cornell University, Ithaca, NY 14853

[2]Cornell High Energy Synchrotron Source (CHESS), Cornell University, Ithaca, NY 14853

[3]School of Applied and Engineering Physics, Cornell University, Ithaca, NY 14853

[4]Physics Department, Cornell University, Ithaca, NY 14853

[5]Department of Materials Science and Engineering, Cornell University, Ithaca, NY 14853

[6]Kavli Institute at Cornell for Nanoscale Science, Ithaca, NY 14853

*Corresponding Author: david.a.muller@cornell.edu

## Abstract

We describe a hybrid pixel array detector (EMPAD - electron microscope pixel array detector) adapted for use in electron microscope applications, especially as a universal detector for scanning transmission electron microscopy. The 128 × 128 pixel detector consists of a 500 μm thick silicon diode array bump-bonded pixel-by-pixel to an application-specific integrated circuit (ASIC). The in-pixel circuitry provides a 1,000,000:1 dynamic range within a single frame, allowing the direct electron beam to be imaged while still maintaining single electron sensitivity. A 1.1 kHz framing rate enables rapid data collection and minimizes sample drift distortions while scanning. By capturing the entire unsaturated diffraction pattern in scanning mode, one can simultaneously capture bright field, dark field, and phase contrast information, as well as being able to analyze the full scattering distribution, allowing true center of mass imaging. The scattering is recorded on an absolute scale, so that information such as local sample thickness can be directly determined. This paper describes the detector architecture, data acquisition (DAQ) system, and preliminary results from experiments with 80 to 200 keV electron beams.

**Introduction**

Electron microscopy is an important tool for the study of materials either through direct imaging, such as in transmission electron microscopy (TEM), or via observing key relevant regions of the electron scattering, as in scanning transmission electron microscopy (STEM). Historically, TEM has been the purview of imaging detectors, where the information is recorded in a single snapshot. STEM has usually been performed using fixed geometry sensors that monitor one or, at most, a few relevant but fixed regions of the scattering distribution. Dynamic range and frame throughput limitations have severely limited imaging devices for STEM work in the past.

Historically, advances in imaging technology, when applied to electron microscopy, have enabled new possibilities in materials analysis. Emulsion films provided good spatial resolution and contrast but non-linear response, low dynamic range, chemical development and post processing made the imaging plate a better alternative (Zuo, 2000). Although imaging plates require post processing, the readout is digital and they provide good linear response and dynamic range, but with limited signal-to-noise ratio (SNR) (Zuo, 2000). The offline nature of emulsion films and imaging plates posed some limitations on advanced imaging experiments demanding automated and high throughput data acquisition (Jin, et al., 2008). Charge couple devices (CCDs) (Boyle & Smith, 1970) are widely used in such applications. Direct conversion in a CCD has excellent single electron sensitivity but suffers with limited dynamic range as the charge created by only a few primary electrons per pixel saturates the CCD. Scintillator-coupled CCDs reduce the average recorded signal per microscope electron, extending the dynamic range per frame but suffer from limited spatial resolution due to electron and optical scattering in scintillator screens (Fan & Ellisman, 2000; Meyer & Kirkland, 1998; Zuo, 2000).

Recent developments in monolithic active pixel sensors (MAPS) and hybrid pixel array detectors (PADs) (Battaglia, et al., 2009; Caswell, et al., 2009; Fan, et al., 1998; Faruqi, et al., 2003; Faruqi, et al., 2005; McGrouther, et al., 2015; McMullan, et al., 2014; Milazzo, et al., 2005) have made them a high-throughput alternative to CCDs. A recent review comparing the performance of commercial MAPS detectors is given by (McMullan, et al., 2014). The MAPS systems include sensor and readout electronics in a single layer, while the PADs usually consist of a thick sensor layer (silicon or CdTe) bump-bonded to the silicon readout electronics underneath. PADs combine the efficiency of a fully-depleted diode sensor coupled pixel-by-pixel to a CMOS readout chip which is custom tailored to the imaging problem at hand. One class of PADs employ a pulse counting architecture which yields a very high S/N ratio for single electron hits but limits the maximum rate of incoming electrons per pixel to about $10^6$ Hz. These detectors work well for imaging at low doses, but saturate quickly when exposed to high fluxes common with diffraction patterns. Alternate PAD architectures which measure the integrated charge do not need to distinguish individual pulses and can extend the rate limitation many orders of magnitude higher. In this architecture, one could choose an integration stage with either a high gain for good single electron sensitivity, or a low gain for a high dynamic range per frame, but not both. A third class of PAD architecture, specifically the electron microscope pixel array detector or "EMPAD" implementation described in this paper, uses a high-gain charge integration front end combined with in-pixel logic to effectively reset the pixel while integration is ongoing. A concurrent trigger to an in-pixel counter extends the dynamic range per frame while the high gain amplifier maintains single electron sensitivity. Some of the high rate capability of charge integrating PADs is sacrificed, but it remains several orders of magnitude higher than pulse counting architectures.

With these high-throughput, high-dynamic range imagers one can now envisage that imaging detectors can be routinely applied to STEM applications. While previous PAD designs have been optimized for real-space imaging, where intensity distributions are relatively uniform and a large number of detector pixels are required, recording diffraction patterns for STEM imaging requires both a high dynamic range and good single-electron sensitivity, which the current design is optimized for, while a high pixel count is less critical. Since the entire scattering image is recorded at each scan point, one can process the data set to simultaneously extract bright field, dark field, center of mass and differential phase contrast information by defining appropriate regions of interest from the detector. However, much more information is contained within the diffraction patterns than is extracted in the typical bright or dark field analyses. Broadening the set of tools used to fully utilize the scattering images is where the true advantages in applying imaging detectors to STEM will become apparent. Some of these ideas have already been tested in earlier detector designs with the limitations discussed above (Cowley, 1993; Kimoto & Ishizuka, 2011; Ozdol, et al., 2015; Zaluzec, 2002). For instance, ptychography for phase recovery that has been demonstrated with CCD detectors (Pennycook, et al., 2015) would benefit from the improved contrast and dynamic range of an EMPAD, as would strain mapping based on nano-beam electron diffraction (Ozdol, et al., 2015).

**Materials and Methods**

**Pixel Array Detector Description:**

A pixel array detector (PAD) consists of a pixelated sensor bump-bonded, pixel-by-pixel, to an underlying application specific integrated circuit (ASIC) which processes the charge generated in each sensor pixel. A typical arrangement of a PAD is shown in Figure 1. Here the sensor is a 500 µm thick diode array fabricated in high-resistivity silicon to allow for full depletion of the diode with a bias of 150 V (SINTEF, Norway). The ASIC is designed and fabricated using standard CMOS tools in a 0.25 µm process (TSMC, Taiwan). Both layers were originally designed as part of a high-dynamic-range, 128 × 128 pixel x-ray imager known as the mixed-mode pixel array detector (MM-PAD) (Angello, 2004; Schuette, 2008; Tate, et al., 2013; Vernon, et al., 2007). This same design turns out to be compatible with electron imaging over a wide primary beam energy range as well. We have tested it for close to a year at up to 200 keV without any obvious degradation in performance.

The pixel design, shown in Figure 2, consists of a front-end amplifier that integrates charge coming from the reverse-biased sensor diode onto a feedback capacitor, $C_{int}$. The output, $V_{out}$, of the integrator is compared with a pre-defined threshold voltage, $V_{th}$. The comparator triggers both a charge removal circuit and an 18-bit, in-pixel counter each time the $V_{out}$ exceeds $V_{th}$ during integration. Each charge removal step takes away a calibrated charge, $\Delta Q$, keeping the input amplifier within its operation range. At the end of integration period, the 18-bit digital value from the counter and the residual analog voltage are read. A field programmable gate array (FPGA) applies a calibrated scale factor to the 18-bit digital value and combines the result with the 12-bit digitization of the analog residual to provide a 30-bit output to the acquisition computer. The value of $C_{int}$ was chosen to be 50 fF for this device to provide high gain for single 8 keV X-ray sensitivity. A future detector designed specifically for electron imaging could easily use a larger integration capacitor to increase the maximum rate of incoming electrons while still maintaining single electron sensitivity.

For readout, the 128 × 128 pixel array is divided into 8 banks of 128 × 16 pixels each. Banks are read in parallel, with pixel outputs buffered sequentially through 8 analog to digital converters and 8 digital I/O lines for the in-pixel counter data. The FPGA is used to control exposure and to receive analog and digital data from the detector. The entire detector can frame continuously at up to 1.1 kHz.

**Detector Assembly**

Figure 3 shows a conceptual design sketch of the prototype detector assembly. The system is composed of three main components – the detector housing on the microscope, the detector control unit (DCU), and a data acquisition and control PC.

**Detector Housing:**

The detector housing is adapted from a Fischione-3000 HAADF detector, with a custom insert containing the EMPAD camera (Figure 4a) replacing the HAADF camera. The Fischione housing pneumatically slides the EMPAD camera insert in and out of the microscope column in order to move the detector in and out of the main electron beam. The insert includes the detector module (Figure 4c), the vacuum feed-through board (Figure 4b), and cooling mechanisms for the detector. The detector module consists of a small printed circuit board which is mounted onto an aluminum heat sink and wire-bonded to the detector ASIC. A long printed circuit board epoxied into a vacuum flange provides the vacuum feedthrough, along with 8 channels of analog to digital conversion and signal buffering to the FPGA.

The detector module is actively cooled by a thermoelectric module to -16.0 +/- 0.05 C to reduce dark current noise and provide thermal stability (the exact temperature is not as important as the temperature stability). The thermoelectric is mounted between the detector module and the vacuum flange, which doubles as a water manifold to provide a heat sink for the thermoelectric module. A copper aperture shields everything but the detector sensor from direct illumination by electrons. Radiation shielding was integrated into the insert to mesh with the shielding present in the Fischione housing to eliminate stray radiation escaping from the microscope.

**Detector Control Unit:**

The detector control unit is connected to the detector housing via a 0.5 m long high-signal density cable to provide control signals to, and data reception from the detector housing. Detector housing power is also provided from this unit, as are signals sent to the external scan coil control of the microscope.

An FPGA board (Xilinx Virtex-6 ML605) controls data acquisition and the communication between different components of the detector system. It drives the detector control signals and captures data from 8 banks of pixels. The captured data is processed and sent to the DAQ computer through a dedicated CameraLink interface. It receives user commands from the DAQ computer using Ethernet communication. Integration timing is set using a 36-bit programmable counter in the FPGA.

The scan control consists of two 16-bit digital-to-analog converters (Texas Instruments DAC8718) which are programmed by the FPGA via high-speed serial peripheral interface (SPI) in synchrony with the image acquisition. The number of scan points and the scan range are programmable. The scan control output is connected to the external scan control input of the microscope.

The DCU includes a custom power supply board that provides regulation and monitoring for different power lines going to the detector chip. The regulation and monitoring is controlled by the FPGA via a dedicated serial SPI interface.

**Data Acquisition and Control Computer:**

The data acquisition computer is a Linux workstation running Ubuntu 12.04 with 64 GB of RAM for image buffering. Images are acquired using a Matrox Radient eCL CameraLink board which can collect frames with a bandwidth of up to 5 Gb/s.

Control is provided with a custom software package which meshes the needed FPGA command sequences with the frame grabbing. The software provides a live display of the diffraction frames as well as the scan images acquired as the electron beam is progressively scanned over the sample. The intensity at each scan image point is computed from user-defined regions of interest (ROIs). Multiple live ROIs can be defined and can easily be set for bright-field, dark-field, or differential phase contrast imaging. A mode which provides continuous imaging at a fixed scan position can be used for both beam and sample alignment. A mode which uses continuous electron beam scanning with a reduced number of scan points provides a refresh rate which is fast enough to allow for microscope focusing and alignment. Before each data session, a dark reference image is determined by averaging at least 100 images taken with no illumination. This reference image is subtracted from each data image coming from the detector and is stable over the course of at least one day. Background subtracted diffraction images from a scan are stored in a single file in a raw block format of 32 bit floating point numbers. A 256 × 256 point scan then results in a file 4 GB in size, and 512 × 512 is 17 GB.

**Detector Characterization:**

To characterize the detector response, a number of imaging tests were performed at various electron energies. The results are summarized in Table 1.

The signal per incident microscope electron was characterized at various electron energies by broadly illuminating the detector at low fluence (<<1 electron/pixel/frame). Each electron of a given energy produces a well-defined number of electron-hole pairs in the sensor (one electron-hole pair for each 3.6 eV of incident microscope electron energy). A histogram of the pixel response over a large number of frames yields an energy response histogram to these monochromatic electrons. Figure 5 shows the pixel response to electrons with 80, 120, 160, and 200 keV energy over 10,000 frames with a mean illumination intensity level of ~ 0.06 electrons/pixel/frame. The curve at each energy has a peak at 0 ADU, corresponding to pixels with no electrons. For the 80 keV curve, discrete peaks at 151, 303, and 454 ADU, correspond to 1, 2, and 3 electrons contained in a single pixel. The curves at higher electron energy have discrete electron peaks at proportionately higher ADU values. These curves together show a response of 1.97 ADU/keV of microscope electron energy. A read noise of 2.8 ADU rms can be computed from the width of the 0 electron peak. This corresponds to a noise of 1.4 keV equivalent microscope electron energy, or 0.014 of a single 100 keV electron. Note that the sensor is constructed with a p+ implant region at the input face that is insensitive to the collection of ionizing radiation. As the incoming electrons traverse this region, a fraction of the energy from each electron will be lost. Thus we observe, for instance, that the 2 electron peak position for 80 keV electrons does not exactly match the position of the 1 electron peak from the 160 keV curve. Fitting to the electron peak positions from the series of energies allows us to estimate a dead region of ~ 1 μm thickness, consistent with parameters obtained from the sensor manufacturer.

The sensor generates a dark current in the absence of illumination at a rate of $7 \times 10^4$ electron-hole pairs/s/pixel at -16 C. This is an equivalent signal to 2.7 primary electrons (at 100 keV) incident on a pixel per second. The noise associated with this dark charge is only 268 electron-hole pairs, equivalent to 0.01 incident electrons (100 keV equivalent) for exposures of up to one second. The dark current is insignificant at typical frame rates of 0.86ms/image.

For 100 keV electrons, the analog well of the integration stage is chosen to accumulate up to 14.2 electrons before the charge removal circuitry is activated. Note that it is not required that an integer number of primary microscope electrons be contained in the analog well. Combining the analog well depth with the 18 bit counter yields a full well of $3.7 \times 10^6$ 100 keV electrons/pixel/frame. Note the charge removal circuitry is limited to a rate of ~ 2MHz, which gives a maximum rate of $2.8 \times 10^7$ 100 keV electrons/pixel/s. Thus for 1 ms images, there is a limit of $2.8 \times 10^4$ 100 keV electrons/pixel/frame. This selected rate corresponds to a beam current of 4.5 pA/pixel. This is not a fundamental limit, but a design choice, and in future designs could be increased by changing the accumulation threshold or counter speed.

The pixel response histograms also show nonzero intensity between the discrete electron peaks due to sharing of charge between pixels. One expects some charge sharing, especially for electrons that strike the sensor near a pixel border. By comparing the area of the first electron peak to that of the charge sharing tail, we can obtain a measure of the charge sharing for randomly placed electron events. Table 2 shows the fraction of incident electrons whose energy is recorded in a single pixel. Note that for higher energies, a higher fraction of the incident electrons have charge shared among pixels, indicating a larger charge spread for these electrons. Monte Carlo simulations were used to estimate the radial distribution of energy deposition in silicon (Win X-ray) (Gauvin, et al., 2006). Table 2 shows the radii at which 50 % and 95 % of the energy from incident electrons is collected. Note that for 200 keV, the radius for complete charge collection is larger than could be contained within a single pixel. If each electron lost energy according to this mean response profile, then no single electron peak would be seen for 200 keV. However, each electron undergoes a random walk through the sensor, losing energy along a unique track, some of which are entirely contained within a pixel. Figure 6 shows 200 such tracks for 80 through 200 keV incident electrons obtained from Win X-ray Monte Carlo simulations.

While single electron tracks can be sharper than the mean spatial response, the spatial response at higher levels of illumination should match the mean response of the sensor. To measure the base response of the system, the knife edge response was measured under illumination by 8 keV x-rays. By using 8 keV x-rays, the energy deposition in the sensor is highly localized to a small volume rather than along a track as with high-energy electrons. Thus the edge response measures the diffusive spread of the charge as it traverses the sensor and is collected at the pixelated nodes. At 150 V sensor bias, this gives a value of 10 - 15 μm FWHM for the diffusive charge spread cloud in the sensor. Under illumination with electrons, this diffusive spread will be convoluted by the charge spread due to the random walk of the primary electron as it loses energy in the sensor. The 50% charge collection radius in Table 2 shows that the width of this convolution should be dominated by the electron track spreading for all but the lowest energies. Measurements of the detector's point spread function and modulation transfer function are given below in the results section (Quantification of Detector Performance). If one were to construct a PAD with smaller pixels, some gain in spatial resolution can be expected for lower electron energies. Little would be gained by using smaller pixels for energies 200 keV or above, unless the detector was thinned as well, which would then introduce straggle into the energy distribution (see discussion of detector optimization below).

**Results and Discussion**

*Imaging examples:*

To investigate the performance of the detector, we mounted it in the 35 mm port of a FEI Tecnai F20 200 keV Schottky field emission STEM. The microscope is uncorrected, and can form a sub-0.2nm spot size with 10 pA of beam current at a 10 mrad convergence semi-angle. The first sample selected was a film of $BiFeO_3$. This is a ferroelectric material grown epitaxially on a $SrRuO_3$ electrode lattice matched to a $DyScO_3$ substrate.

As an illustration of the detectors sensitivity and dynamic range, Figure 7 shows the diffraction pattern recorded along the [010] pseudocubic zone axis in $BiFeO_3$. Figure 7a shows a pattern recorded in 1 ms and displayed on a log scale, calibrated in the number of primary electrons. Even at 1 ms, the recorded diffraction pattern shows both the central beam and Kikuchi band details out to the HOLZ line, where the high SNR ratio for single electron detection is essential for resolving the HOLZ line itself. Fig 7b shows the accumulation of data over 100 frames. The data now spans over 4 orders of magnitude and the details of the unsaturated central beams as well as the Kikuchi bands and HOLZ lines are much clearer.

Throughout our work, we have not yet been able to saturate the detector, even with the unscattered beam placed directly on the detector. This allows us to record full diffraction patterns without the need for a beam stop. Consequently, we can quantify any extracted signal as either a fraction of the incident beam, or as an absolute number of electrons. The former is useful for quantitative atom counting and thickness determination (LeBeau, et al., 2010), and avoids the challenges of having to correct for the integrated, non-uniform response of an ADF scintillator.

With these issues in mind, Figure 8 shows the bright field (BF), High Angle Annular Dark Field (HAADF), Differential Phase Contrast (DPC), and Center of Mass (COM) signals all extracted from the same EMPAD-STEM data set. All signals are therefore aligned to each other, and their relative (and absolute) intensities can be compared directly. The HAADF signal from 50 to 250 mrad (Fig 8b) is plotted as a fraction of the incident beam ($I/I_0$), which can be compared to multislice simulations to extract the local variations in thickness. Even signals whose overlapping nature would make them incompatible if recorded on distinct detectors (e.g., BF (a) or annular bright field versus DPC) can be simultaneously determined. The bright field image from 0 to 5mrad is shown in Fig 8a. The DPC and COM signals in Fig 8c,d require integration over the entire diffraction pattern to meet the strong quantum measurement condition (Lubk & Zweck, 2015), and ensure lateral incoherence (Majert & Kohl, 2015; Rose, 1976; Waddell & Chapman, 1979).

For very thin specimens, the net deflection of the electron beam can be related to the electric field in the sample (Chapman, et al., 1978; Kimoto & Ishizuka, 2011; Lubk & Zweck, 2015; Muller, et al., 2014; Rose, 1976; Waddell & Chapman, 1979). For thicker objects or imaging at atomic resolution, considerably more caution is required for the interpretation of these deflection images (Lubk & Zweck, 2015; MacLaren, et al., 2015). From a theoretical standpoint, calculating the center of mass of the diffraction pattern, $I(\vec{p})$, gives

$$\langle \vec{p} \rangle = \int \vec{p} I(\vec{p}) d\vec{p}, \tag{1}$$

the expectation value of the net momentum transfer, $\langle \vec{p} \rangle$, integrated through the depth of the film (Lubk & Zweck, 2015; Muller, et al., 2014; Waddell & Chapman, 1979). With previous detectors, direct measurement and integration of the diffraction pattern has not been possible, making Figure 8d probably the first experimental realization of direct center of mass imaging. Instead, the approach has been to employ a split or quadrant detector (Chapman, et al., 1978; Dekkers & Lang, 1974; Rose, 1974) to measure the deflection of the beam by shifts of intensity between the split segments. The resulting differential phase contrast (DPC) image provides a qualitative approximation to the COM image – the DPC image of figure 8c is visually similar, albeit slightly noisier, to the COM image of figure 8d. More careful analysis shows that the DPC phase contrast transfer function is anisotropic, with about 20-30% reduction in transfer at midrange frequencies, and the amplitude contrast transfer function has a cutoff along $k_x$ at about half that of the COM detector(Waddell & Chapman, 1979).

With the EMPAD, both the DPC and COM signals are easy to extract. The convention of normalizing the DPC signal as the difference/sum of the split segments to minimize diffraction contrast does introduce nonlinearities and the potential for numerical instabilities as is evident at the top of Figure 8c. In terms of noise, the COM signal appears to perform more robustly in our experience (contrast with Figure 8d). As a matter of convenience, the COM signal has one further advantage, that is once the diffraction pattern (such as Figure 7) is calibrated, then it follows from equation (1) that so the COM image is also calibrated. In general, once we know the camera length, the same calibration can be used to a few percent accuracy for the COM image. Quantification of the fractional DPC shift is more problematic, and depends on knowing the shape of the diffraction, with small changes in shape and centering leading to artifacts (Takahashi & Yajima, 1994; Yajima, 2009).

Atomic-resolution imaging is also possible with the EMPAD detector. Figure 9 shows the HAADF and $x$ and $y$ COM images of a domain boundary in the $BiFeO_3$ film. While there is a clear shift of the beam between the two domains, it would be incorrect to interpret this as an electric field, consistent with recent cautions (MacLaren, et al., 2015; Waddell & Chapman, 1979). In both Figures 8 and 9, the domains in the $BiFeO_3$ film are oriented along the [111] zone, while the film is viewed down the [001] pseudocubic zone. The distortion of the unit cell in the ferroelectric state forces a tilt of the domains to retain coherent boundaries, with a resulting tilt of the diffraction pattern on the order of mrad. By examining EMPAD-STEM images with a smaller convergence angle, the tilting of the Ewald sphere across these domains is very obvious. The contrast in the COM images is dominated by a mixture of tilt and polarity difference. In these images, tilt plays a larger role than polarity, but that will depend on the sample geometry. The large integration angles appear to suppress any Fresnel contrast at the boundary in general, although these particular images are recorded in focus. We also note contrast differences with thickness as diffracted beams oscillate vary in intensity with thickness.

A final imaging example of true field measurement is shown in Figure 10. Here a 50 nm-thick Co film is imaged in low-magnification STEM (LM-STEM) mode with the objective lens turned off for field-free imaging. In this configuration the beam is highly parallel, which makes COM detection highly sensitive to small shifts in magnetic field. Additionally, the beam is larger than the grain size in the cobalt film, thus averaging over the grain contrast that is apparent in regular STEM mode with a smaller spot size. The deflection of the beam appears as a uniform shift of a few microradians. With the film thickness known, the deflection is converted to magnetic field – e.g., equation 1 of (Chapman, 1984). The result is a quantitative image of both the x and y components of the magnetic field ripples in the cobalt film. Again, the center of mass analysis gives a simple and direct calibrated image.

*Quantification of Detector Performance:*

To investigate the spatial response of the detector, we measured the line spread function (LSF) and the modulation transfer function (MTF) from an edge resolution test. We imaged the edge of a large (800 micron) selected area aperture projected onto the detector at low magnification in TEM mode. We summed 100 successive images of the edge (0.1 s total acquisition time), and calculated the LSF from the derivative of the image. The results are shown in Figure 11 a,b for 80 and 200 keV respectively. We calculated the MTFs from the Fourier transform of the LSF and these are shown in Figure 11c. The effect of the increased LSF tails at 200 keV vs 80 keV is apparent, and is expected, given the spread seen in the simulations of figure 6.

To measure the detective quantum efficiency (DQE) we use the noise binning method (McMullan, et al., 2009) and extrapolate the measured noise as a function of bin size to account for the effect of the point spread function. We work from a series of 100 images and use the difference between successive images to follow the same approach as has been used to benchmark previous MAPS detectors (McMullan, et al., 2009). The DQE is a function of dose rate – at low rates readout and dark noise will dominate, and at high rates gain variations between pixels become significant. For a count rate of roughly 0.2 electrons/pixel/frame, the extrapolated DQE at 80 kV is 0.88, and at 200 keV is 0.93. For a mid range count rate of roughly 700 electrons/pixel/frame, the extrapolated DQE at 80 kV is 0.93, and at 200 keV is 0.94. The difference is probably within the extrapolation error.

The DQE for systems that operate in counting mode, are typically reported at doses of well below 1 electron/pixel as the DQE of counting mode systems degrades as the count rate increases due to coincidence losses (Appendix C of (McMullan, et al., 2014) suggests roughly a 10% loss in DQE at 10 electrons/pixel/frame). The EMPAD, which relies on charge integration instead, retains a high DQE over many orders of magnitude more than that.

While a frequency dependent DQE is often used to characterize imaging detectors that operate with relatively flat fields such as imaging low-contrast objects, it does less well at capturing the key performance properties of a detector intended for diffraction work. Diffraction patterns display very large dynamic ranges, and spill over from tails of high intensity diffraction peaks can swamp small features nearby. A similar challenge is present for detectors used for electron energy loss spectroscopy when measuring features near the zero loss peak. With these larger intensity variations, measures such as the full width at $1/100^{th}$ maximum (FWCM) and full width at $1/1000^{th}$ maximum (FWKM) are a better reflection of diffraction performance. For the EMPAD, the FWCM is 3 pixels at both 80 and 200 keV, and the FWKM is 3 pixels at 80 keV and 5 pixels at 200 keV. A comparison measurement of the EMPAD's FWKM with other reported direct electron detectors to date does not appear possible as the signal required to measure a $1/1000^{th}$ of the maximum exceeds their designed dynamic ranges per frame.

**Discussion of Detector Optimization**

There are two main aspects of the detector where trade-offs can be considered: first, how many pixels do we need for a diffraction camera, and second, how many electrons per pixels should we be able to record? To a fair extent, these two questions are coupled. A thick sensor will require large pixels to match the inherent charge spread, but a thicker sensor will also collect electrons far more efficiently. For imaging sensors, a high pixel count is desirable, but as a STEM detector, extra sensor pixels are undesirable because of resulting bandwidth and storage problems for high speed readouts. Moving from a 128×128 pixel sensor to a 256×256 pixel sensor will quadruple the storage requirements per 4D EMPAD-STEM map from 4GB to 16 GB for a simple 256×256 probe position map.

The question of how many sensor element pixels are needed will depend on the measurement planned. For phase reconstruction of the bright field disk using ptychography, there is a diminishing return beyond 16×16 pixel elements for an atomic-sized probe (Yang, et al., 2015). For COM and its extreme limit of DPC imaging, obviously it is possible to detect very small beam deflections using only 2×2 pixels. The question of what is the optimum number of pixels has been studied heavily in the optical super-resolution community where the location of fluorescent centers to sub wavelength accuracy is essential. Thompson, Larson and Webb (Thompson, et al., 2002), have shown that the optimal ratio of sensor size, $a$, to spot size, $s$, is

$$(a/s)^4 = 96\pi b^2/N , \qquad \text{-(2)}$$

where N is the number of primary counts in the beam and $b$ is noise/pixel in primary counts. Here $s$ is the standard deviation of the point spread function, which is assumed to be Gaussian. A similar analysis with a different scaling coefficient would hold for a top hat function. The main trend is that for a noisy detector, fewer elements are desirable, and as the detector's signal/noise ratio improves, more detector elements can be considered. For instance, for our detector at 200 keV with N=1000 primary electrons, 16 pixels would be optimal for locating the center of mass. At 10,000 electrons, this grows to 28 pixels, with an accuracy of 1% of the spot size, although the curve is relatively flat vs number of pixels, with the root mean square error given by (Thompson, et al., 2002)

$$\left\langle (\Delta x)^2 \right\rangle = \frac{s^2 + a^2/12}{N} + \frac{4\sqrt{\pi}s^2 b^2}{aN^2} \qquad -(3).$$

This analysis is worth keeping in mind as we consider more detailed analyses of diffraction patterns. About 16 pixels per diffraction disk is sufficient for tracking the center, thus our 128×128 pixel sensor could record 8×8 diffraction spots in parallel if optimally configured. For differentiating between polarity and tilt contributions to the COM signal, imaging diffraction patterns with non-overlapping disks can be very helpful.

For measuring the shift of excess and deficit HOLZ lines, more pixels might be desirable, but the contrast also depends on the number of electrons per pixel, so large signals and beam are required. Many of the monolithic active pixel sensors (MAPS) use relatively thin active layers and as a result can only hold relatively few (25 to 1000 electrons) primary electrons per pixel, severely limiting their dynamic range, and the angular range that they can record simultaneously. In exchange, the thin layer results in less charge spreading, and as a result can use a smaller pixel size. Our current EMPAD design uses larger and deeper active pixel volumes (150×150×500 microns) which greatly improves the SNR/primary electron, reducing $b$ in equations 2 and 3.

One benefit of the thick detector where almost all the energy from each electron is collected by the detector is that the energy distribution is relatively narrow. In contrast, thin transmission detectors which are designed to allow most electrons to pass through with little spatial spread, result in an energy spread that follows a Landau distribution e.g. (McMullan, et al., 2014). The Landau distribution has a large straggle and a mean energy is very sensitive to the energy cutoff. Thus charge integration with a thin detector would result in a very noisy signal that would be difficult to quantify. The thin detectors obtain their good signal to ratios by operating in counting mode instead, resulting in both limited dynamic range and low count rates per pixel, but can use a much smaller pixel size.

The thick and thin detectors have very different energy dependences. The point spread increases with primary voltage for the charge-integrating thick detector, which is the reverse to that found for thin sensors operating in counting mode. At higher voltages, less than 100% of the energy is deposited in the thin sensor, leading to reduced sensitivity but an improved MTF as fewer electrons have an opportunity to scatter within the pixel. In contrast the thick sensor retains a high sensitivity, but exhibits a degraded MTF once the beam spread becomes comparable to the pixel size.

The existing pixel array module was designed for x-ray imaging and no modifications were made to the ASIC or sensor when it was adapted to the present STEM application. Redesign of the pixel array sensor and ASIC are certainly feasible, though by no means trivial. For example, use of high atomic weight sensor materials (e.g., CdTe) could help to increase the pixel count, as per the discussion above, without increasing the detector area or compromising the SNR. Sensors with pixels of a different area and/or thickness are possible, as are ASICs that frame more quickly and have a higher dynamic range. However, it should be noted that pixel array detector design is typically sufficiently expensive (millions of dollars) and time consuming (many person years) that redesign should be approached with a great deal of thought and in response to pressing unfulfilled requirements. Thus, in our opinion, it is likely wisest to gain experience with the present design on many more real materials applications before plunging into a redesign effort of the fundamental MM-PAD module.

**Conclusion**

This high dynamic-range EMPAD provides convenience in defining custom regions of interest which can reproduce any of the conventional STEM imaging metrics such as bright field, annular dark field, or differential phase contrast. However, the real impact of this type of detector will likely be the ability to define new modalities which will allow quantitative measures and decoupling of parameters such as tilt, electric and magnetic fields to be extracted from a data set, or since almost every electron that is transmitted through the sample is recorded, the potential for post-hoc optimally dose-efficient image reconstruction. Recording the full diffraction pattern on one detector allows the scattering to be placed on an absolute scale, with a direct measurement of the magnitude of diffraction spot deflections possible. Careful consideration of effects due to sample tilt and thickness must be given to separate those effects from underlying deflections from internal sample fields. This may require one to choose to image with beam optics to be optimized and aligned with non-standard parameters.

The range of electron energies could be extended well below the 80 keV tested here. As an x-ray imager, good signal to noise for single x-rays has been demonstrated down to 8 keV. Lower energies than this can be used since the charge integration nature does not need to resolve individual incoming quanta. The MM-PAD detector has been used successfully with X-ray energies as low as 2.5 keV. The lower energy threshold is limited primarily by the aluminized bias contact and the diode p+ implant on the input side of the sensor. At higher electron energies (over 200 keV), one must consider lattice displacement damage in the sensor, and ultimately, radiation damage to the readout chip once the electron range in silicon becomes greater than 500 μm. Alternative, higher atomic number sensor materials could alleviate these concerns to some degree, as well as improve the spatial resolution of the detector.

While performance of this converted x-ray imager has been quite good for imaging electron scattering, some improvements could improve the utility of the device for use on electron microscopes. The addition of a data latch in the pixel would allow integration during readout, which would improve the duty cycle of operation. Secondly, since the signal/noise ratio is very high for electrons in this energy range, one could implement amplifiers with higher bandwidth (and higher noise) to reduce read time and improve frame rate. At higher frame rate, an improvement to the incoming electron rate would prove beneficial. One can choose a larger integration capacitor and optimize the charge removal circuitry to improve the maximum incoming rate. Changes to the readout architecture such as implementing selective addressing of pixels or selective ganging of pixels could also improve read speed to 10 kHz or beyond.

With the present detector format, there is an inherent trade-off in choosing the camera field of view for diffraction imaging. Capturing high-angle scattering usually used for HAADF images will limit the angular resolution at which details of lower-order diffraction disks are acquired. A larger format detector can alleviate this. Besides the higher data collection bandwidth and storage needed, the maximum size of the ASIC die is limited. Larger x-ray detectors of similar design have already been constructed using multiple ASICs attached to a larger sensor.

**Acknowledgements**

PAD development in SMG's lab is supported by the U.S. Department of Energy, grant DE-FG02-10ER46693 and by CHESS, which is supported by the U.S. National Science Foundation and the U.S. National Institutes of Health-National Institute of General Medical Sciences via grant DMR-1332208. The adaptation to the STEM was supported by the Kavli Institute at Cornell for Nanoscale Science. Data acquisition (K.X.N., D.A.M.) was supported by the Cornell Center for Materials Research, an NSF MRSEC, grant DMR 1120296. We thank Deyang Chen and Ramamoorthy Ramesh for providing us with the $BiFeO_3$ sample and Paul Fischione for providing the base ADF housing.

**References**

Angello, A.G., Augustine, F., Ercan, A., Gruner, S., Hamlin, R., Hontz, T., Renzi, M., Schuette, D., Tate, M., Vernon, W. (2004). Development of a mixed-mode pixel array detector for macromolecular crystallography. IEEE Nucl. Sci. Symposium **7**, 4667-4671.

Battaglia, M., Contarato, D., Denes, P., Doering, D., Giubilato, P., Kim, T.S., Mattiazzo, S., Radmilovic, V. & Zalusky, S. (2009). A rad-hard CMOS active pixel sensor for electron microscopy. Nuclear Instruments and Methods in Physics Research Section A: Accelerators, Spectrometers, Detectors and Associated Equipment **598**(2), 642-649.

Boyle, W.S. & Smith, G.E. (1970). Charge Coupled Semiconductor Devices. Bell System Technical Journal **49**(4), 587.

Caswell, T.A., Ercius, P., Tate, M.W., Ercan, A., Gruner, S.M. & Muller, D.A. (2009). A high-speed area detector for novel imaging techniques in a scanning transmission electron microscope. Ultramicroscopy **109**(4), 304-311.

Chapman, J.N. (1984). The Investigation Of Magnetic Domain-Structures In Thin Foils By Electron-Microscopy. Journal of Physics D-Applied Physics **17**(4), 623-647.

Chapman, J.N., Batson, P.E., Waddell, E.M. & Ferrier, R.P. (1978). The direct determination of magnetic domain wall profiles by differential phase contrast electron microscopy. Ultramicroscopy **3**(0), 203-214.

Cowley, J.M. (1993). Configured detectors for STEM imaging of thin specimens. Ultramicroscopy **49**(1–4), 4-13.

Dekkers, N.H. & Lang, H.D. (1974). Differential Phase-Contrast In A STEM. Optik **41**(4), 452-456.

Fan, G.Y., Datte, P., Beuville, E., Beche, J.F., Millaud, J., Downing, K.H., Burkard, F.T., Ellisman, M.H. & Xuong, N.H. (1998). ASIC-based event-driven 2D digital electron counter for TEM imaging. Ultramicroscopy **70**(3), 107-113.

Fan, G.Y. & Ellisman, M.H. (2000). Digital imaging in transmission electron microscopy. Journal of Microscopy-Oxford **200**, 1-13.

Faruqi, A.R., Cattermole, D.M., Henderson, R., Mikulec, B. & Raeburn, C. (2003). Evaluation of a hybrid pixel detector for electron microscopy. Ultramicroscopy **94**(3-4), 263-276.

Faruqi, A.R., Henderson, R. & Tlustos, L. (2005). Noiseless direct detection of electrons in Medipix2 for electron microscopy. Nuclear Instruments & Methods in Physics Research Section a-Accelerators Spectrometers Detectors and Associated Equipment **546**(1-2), 160-163.

Gauvin, R., Lifshin, E., Demers, H., Horny, P. & Campbell, H. (2006). Win X-ray: A New Monte Carlo Program that Computes X-ray Spectra Obtained with a Scanning Electron Microscope. Microscopy and Microanalysis **12**(01), 49-64.

Jin, L., Milazzo, A.C., Kleinfelder, S., Li, S.D., Leblanc, P., Duttweiler, F., Bouwer, J.C., Peltier, S.T., Ellisman, M.H. & Xuong, N.H. (2008). Applications of direct detection device in transmission electron microscopy. Journal of Structural Biology **161**(3), 352-358.

Kimoto, K. & Ishizuka, K. (2011). Spatially resolved diffractometry with atomic-column resolution. Ultramicroscopy **111**(8), 1111-1116.

LeBeau, J.M., Findlay, S.D., Allen, L.J. & Stemmer, S. (2010). Standardless Atom Counting in Scanning Transmission Electron Microscopy. Nano Letters **10**(11), 4405-4408.

Lubk, A. & Zweck, J. (2015). Differential phase contrast: An integral perspective. Physical Review A **91**(2), 023805.

MacLaren, I., Wang, L.Q., McGrouther, D., Craven, A.J., McVitie, S., Schierholz, R., Kovacs, A., Barthel, J. & Dunin-Borkowski, R.E. (2015). On the origin of differential phase contrast at a locally charged and globally charge-compensated domain boundary in a polar-ordered material. Ultramicroscopy **154**, 57-63.

Majert, S. & Kohl, H. (2015). High-resolution STEM imaging with a quadrant detector Conditions for differential phase contrast microscopy in the weak phase object approximation. Ultramicroscopy **148**, 81-86.

McGrouther, D., Krajnak, M., MacLaren, I., Maneuski, D., O'Shea, V. & Nellist, P.D. (2015). Use of a hybrid silicon pixel (Medipix) detector as a STEM detector. Microscopy and Microanalysis **21**(Supplement S3), 1595-1596.

McMullan, G., Chen, S., Henderson, R. & Faruqi, A.R. (2009). Detective quantum efficiency of electron area detectors in electron microscopy. Ultramicroscopy **109**(9), 1126-1143.

McMullan, G., Faruqi, A.R., Clare, D. & Henderson, R. (2014). Comparison of optimal performance at 300 keV of three direct electron detectors for use in low dose electron microscopy. Ultramicroscopy **147**, 156-163.

Meyer, R.R. & Kirkland, A. (1998). The effects of electron and photon scattering on signal and noise transfer properties of scintillators in CCD cameras used for electron detection. Ultramicroscopy **75**(1), 23-33.

Milazzo, A.-C., Leblanc, P., Duttweiler, F., Jin, L., Bouwer, J.C., Peltier, S., Ellisman, M., Bieser, F., Matis, H.S., Wieman, H., Denes, P., Kleinfelder, S. & Xuong, N.-H. (2005). Active pixel sensor array as a detector for electron microscopy. Ultramicroscopy **104**(2), 152-159.

Muller, K., Krause, F.F., Beche, A., Schowalter, M., Galioit, V., Loffler, S., Verbeeck, J., Zweck, J., Schattschneider, P. & Rosenauer, A. (2014). Atomic electric fields revealed by a quantum mechanical approach to electron picodiffraction. Nature Communications **5**, 5653.1-6.

Ozdol, V.B., Gammer, C., Jin, X.G., Ercius, P., Ophus, C., Ciston, J. & Minor, A.M. (2015). Strain mapping at nanometer resolution using advanced nano-beam electron diffraction. Applied Physics Letters **106**(25).

Pennycook, T.J., Lupini, A.R., Yang, H., Murfitt, M.F., Jones, L. & Nellist, P.D. (2015). Efficient phase contrast imaging in STEM using a pixelated detector. Part 1: Experimental demonstration at atomic resolution. Ultramicroscopy **151**, 160-167.

Rose, H. (1974). Phase-Contrast In Scanning-Transmission Electron-Microscopy. Optik **39**(4), 416-436.

Rose, H. (1976). Nonstandard imaging methods in electron microscopy. Ultramicroscopy **2**, 251-267.

Schuette, D.R. (2008). A mixed analog and digital pixel array detector fro synchrotron x-ray imaging. Ph.D. thesis, Cornell University, Ithaca, NY USA. **Ph.D.**

Takahashi, Y. & Yajima, Y. (1994). Nonmagnetic contrast in scanning Lorentz electron microscopy of polycrystalline magnetic films. Journal Of Applied Physics **76**(12), 7677-7681.

Tate, M.W., Chamberlain, D., Green, K.S., Philipp, H.T., Purohit, P., Strohman, C. & Gruner, S.M. (2013). A Medium-Format, Mixed-Mode Pixel Array Detector for Kilohertz X-ray Imaging. 11th International Conference on Synchrotron Radiation Instrumentation (SRI 2012) **425**, 062004.

Thompson, R.E., Larson, D.R. & Webb, W.W. (2002). Precise nanometer localization analysis for individual fluorescent probes. Biophysical Journal **82**(5), 2775-2783.

Vernon, W., Allin, M., Hamlin, R., Hontz, T., Nguyen, D., Augustine, F., Gruner, S.M., Xuong, N.H., Schuette, D.R., Tate, M.W. & Koerner, L.J. (2007). First results from the 128x128 pixel mixed-mode Si x-ray detector chip - art. no. 67060U. Hard X-Ray and Gamma-Ray Detector Physics IX **6706**, U7060-U7060.

Waddell, E.M. & Chapman, J.N. (1979). Linear Imaging Of Strong Phase Objects Using Asymmetrical Detectors In STEM. Optik **54**(2), 83-96.

Yajima, Y. (2009). Lorentz Scanning Transmission Electron Microscopy (Lorentz STEM): Model Analyses of Detector Performance. Bull. Col. Edu. Ibaraki Univ. (Nat. Sci.) **58** 19-24.

Yang, H., Pennycook, T.J. & Nellist, P.D. (2015). Efficient phase contrast imaging in STEM using a pixelated detector. Part II: Optimisation of imaging conditions. Ultramicroscopy **151**, 232-239.

Zaluzec, N.J. (2002). Quantitative Measurements Of Magnetic Vortices Using Position Resolved Diffraction In Lorentz STEM. Microscopy and Microanalysis **8**(Supplement S02), 376-377.

Zuo, J.M. (2000). Electron detection characteristics of a slow-scan CCD camera, imaging plates and film, and electron image restoration. Microscopy Research and Technique **49**(3), 245-268.

Table 2

| | Table 2. Charge spread in sensor | | |
|---|---|---|---|
| Energy (keV) | Fraction of single pixel electrons | Radius (μm) to collect 50% of charge | Radius (μm) to collect 95% of charge |
| 80 | 0.54 | 14 | 32 |
| 120 | 0.40 | 30 | 67 |
| 160 | 0.26 | 49 | 112 |
| 200 | 0.13 | 74 | 167 |

**Figure Captions**

**Figure 1.** a.) Schematic of STEM imaging using the EMPAD in where the beam is stepped at each scan position and the full CBED diffraction pattern is recorded. b) Schematic of the physical structure of the EMPAD. The pixelated sensor (shown in blue with a corner removed) is bump-bonded (solder bumps represented here as gold balls) pixel-by-pixel to the underlying signal processing chip (in pink) (from Angello, *et al.,* 2004). (The pixel divisions shown on the top of the sensor are conceptual. In practice, pixels are defined the electric field lines determined by metallization on the bump-bond side of the sensor.)

**Figure 2**. Pixel schematic of the mixed-mode PAD. Charge generated in the reversed biased diode at left is integrated onto a capacitor, Cint. A comparator stage compares the integrator voltage, Vout, with a defined threshold, Vth. When Vout drops below Vth during exposure, circuitry is triggered which removes a charge, $\Delta Q$, from the integrator, keeping the integrator in its operating range. At the same time, an in-pixel counter is incremented. At exposure end, Vout is digitized and combined with the digital output of the 18-bit counter to yield a high-dynamic range value for the pixel signal.

**Figure 3.** Detector Overview. The Detector Housing connects to a standard camera vacuum port on the microscope. The Detector Control Unit (DCU) provides control signals to the detector and captures the image data stream. Beam scan voltages are synchronized with image acquisition in SEM mode and are connected to the microscope's external scan control inputs. Images are transferred to the data acquisition computer via a CameraLink interface. Instructions for camera control and setup are sent to the DCU via Ethernet commands.

**Figure 4.** Detector Housing and camera insert. a) Fischione housing with EMPAD detector mounted on microscope. Also shown is the Detector Control Unit (DCU) b) EMPAD camera insert. The EMPAD detector module is at far left and is plugged into a long printed circuit board which is epoxied into a vacuum flange. Items to the left of the flange are in vacuum. Connectors on the right provide power and control signals. Not shown are the copper aperture covering the detector module and the additional radiation shielding near the connectors. c) Detail of detector module. The EMPAD sensor is the small gray square on the left, which is wire-bonded to the dark printed circuit board. These pieces are attached to an aluminum heat-sink. A thermoelectric module (not shown) is sandwiched between this heat-sink and the larger aluminum piece underneath, which is machined as one part with the vacuum flange as acts as a water cooled manifold for the thermoelectric module.

**Figure 5.** Pixel response histogram from 10,000 frames for very low, uniform illumination of the detector with electrons of 80, 120, 160, and 200 keV energy. The peak at 0 ADU corresponds to pixels with no electrons. For the 80 keV curve, the peaks at 151, 303, and 454 ADU correspond to peaks from 1, 2 and 3 microscope electrons per pixel. Intensity between the peaks arises from the signal of some electrons being shared between adjacent pixels. Discrete electron peaks for higher energy electrons occur at proportionately higher ADU values.

**Figure 6.** Monte Carlo simulations of electron tracks in silicon for a) 80, b) 120, c) 160, and d) 200 keV electrons. Each panel shows 200 individual electron tracks which impinge on the sensor at the top of each panel (3 pixel area covered in each panel). Pixel dimensions of 500 μm thickness (top to bottom) and 150 μm laterally are indicated by gray lines.

**Fig 7:** a.) CBED pattern of $BiFeO_3$ recorded in a) 1 ms and b) 100ms with 10 pA of beam current at a single scan position. All CBED images shows the number of primary electrons detected on the EMPAD detector, showing quantitative measurements of electron counting. Black bar on the lowermost far right represents 20 mrad for the diffraction patterns above.

**Fig 8.** Different imaging modes extracted from a EMPAD-STEM image of 137 nm thick $BiFeO_3$ film grown on a 54 nm $SrRuO_3$ electrode on a $DyScO_3$ substrate. The acquisition time was 65 seconds. a.) Bright field signal from 0 to 5 mrad, b) annular dark field signal from 50-250 mrad, c) differential phase contrast in the x-direction and d) center of mass shift in the x-direction. COM deflections are shown in milliradians. The black scale bar under d), which is common to all the images is 50 nm. The $(I/I_0)$ colorbar is common to the BF and HAADF images of panels a and b where is the $I_0$ incident beam intensity. The striations are aliasing artifacts from undersampling of the atomic lattice with an atomic-sized probe.

**Fig 9:** a) Atomic resolution HAADF image of $BiFeO_3$ taken using the EMPAD at a domain boundary. b) center of mass image taken in x and c) y showing structure changes at the atomic domain boundaries where deflections are shown in milliradians. The 2 nm scale bar in c also corresponds to a and b. All images are extracted from the same EMPAD-STEM data set, with 65 second acquisition time.

**Fig 10:** Center of Mass measurements of magnetic deflections (converted from mrad to Tesla) for 50 nm Co specimen on silicon nitride in a) x- and b) y-directions. Black scale bar represents 4 microns. Recorded in LM-STEM mode with the objective lens turned off for field free-imaging.

**Fig 11**: Measured detector response from the edge of an aperture imaged onto the detector. Line spread functions for (a) 80 and (b) 200 keV were calculated from the derivative of 100 images of the edge. The modulation transfer function (c) obtained from the Fourier transforms of the line spread functions, with spatial frequency plotted as a fraction of the Nyquist frequency.

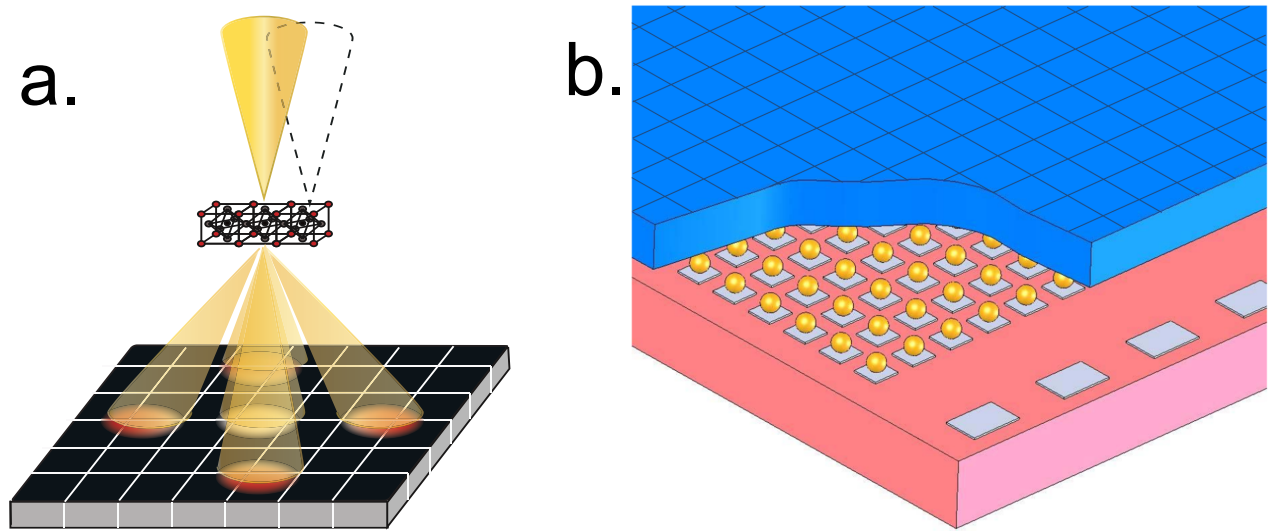

Figure 1. a.) Schematic of STEM imaging using the EMPAD in where the beam is stepped at each scan position and the full CBED diffraction pattern is recorded.  b) Schematic of the physical structure of the EMPAD. The pixelated sensor (shown in blue with a corner removed) is bump-bonded (solder bumps represented here as gold balls) pixel-by-pixel to the underlying signal processing chip (in pink) (from Angello, *et al.,* 2004). (The pixel divisions shown on the top of the sensor are conceptual. In practice, pixels are defined the electric field lines determined by metallization on the bump-bond side of the sensor.)

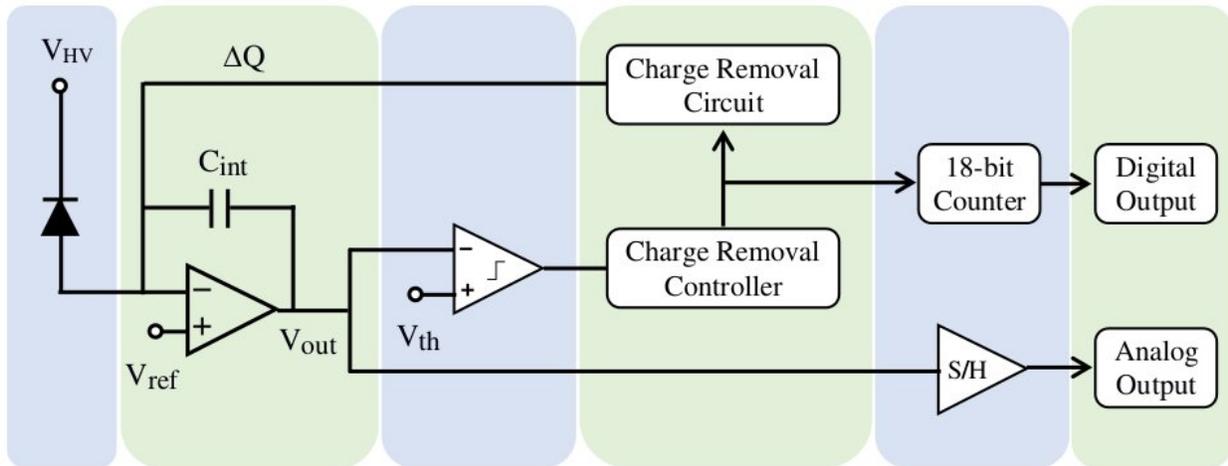

Figure 2. Pixel schematic of the mixed-mode PAD. Charge generated in the reversed biased diode at left is integrated onto a capacitor, Cint. A comparator stage compares the integrator voltage, Vout, with a defined threshold, Vth. When Vout drops below Vth during exposure, circuitry is triggered which removes a charge, ΔQ, from the integrator, keeping the integrator in its operating range. At the same time, an in-pixel counter is incremented. At exposure end, Vout is digitized and combined with the digital output of the 18-bit counter to yield a high-dynamic range value for the pixel signal.

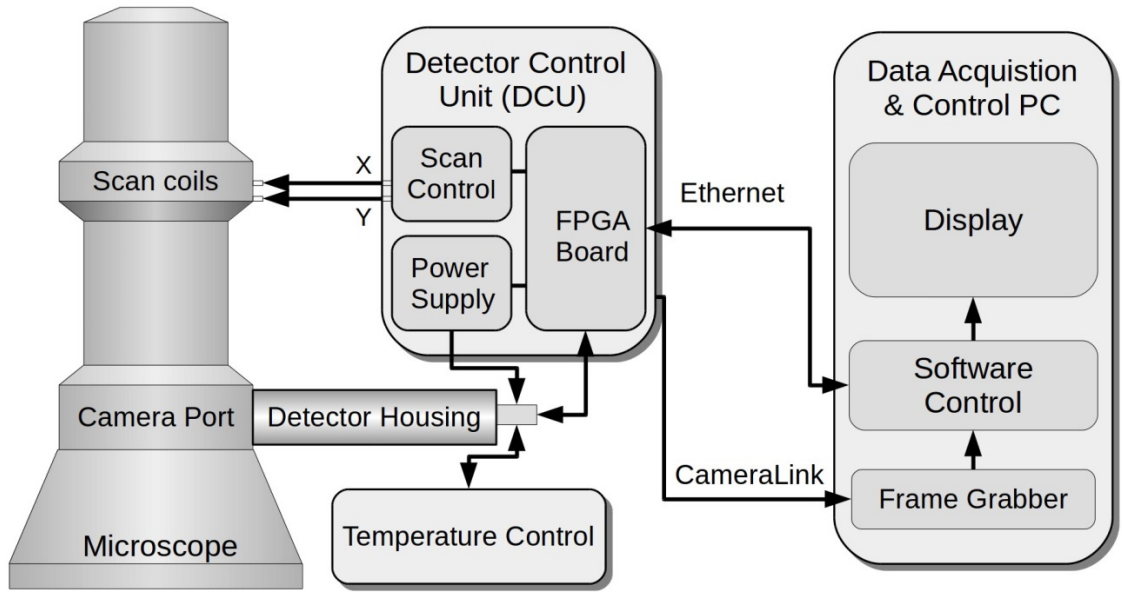

Figure 3. Detector Overview. The Detector Housing connects to a standard camera vacuum port on the microscope. The Detector Control Unit (DCU) provides control signals to the detector and captures the image data stream. Beam scan voltages are synchronized with image acquisition in SEM mode and are connected to the microscope's external scan control inputs. Images are transferred to the data acquisition computer via a CameraLink interface. Instructions for camera control and setup are sent to the DCU via Ethernet commands.

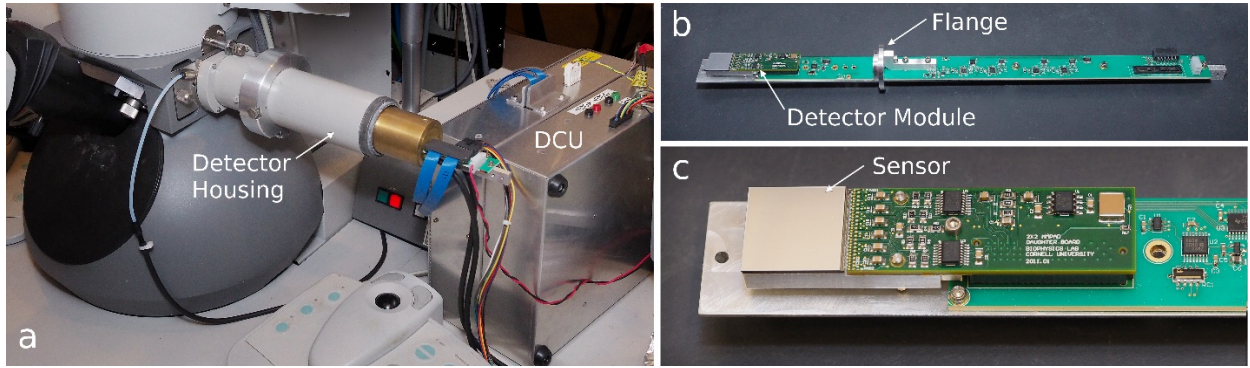

Figure 4. Detector Housing and camera insert. a) Fischione housing with EMPAD detector mounted on microscope. Also shown is the Detector Control Unit (DCU) b) EMPAD camera insert. The EMPAD detector module is at far left and is plugged into a long printed circuit board which is epoxied into a vacuum flange. Items to the left of the flange are in vacuum. Connectors on the right provide power and control signals. Not shown are the copper aperture covering the detector module and the additional radiation shielding near the connectors. c) Detail of detector module. The EMPAD sensor is the small gray square on the left, which is wire-bonded to the dark printed circuit board. These pieces are attached to an aluminum heat-sink. A thermoelectric module (not shown) is sandwiched between this heat-sink and the larger aluminum piece underneath, which is machined as one part with the vacuum flange as acts as a water cooled manifold for the thermoelectric module.

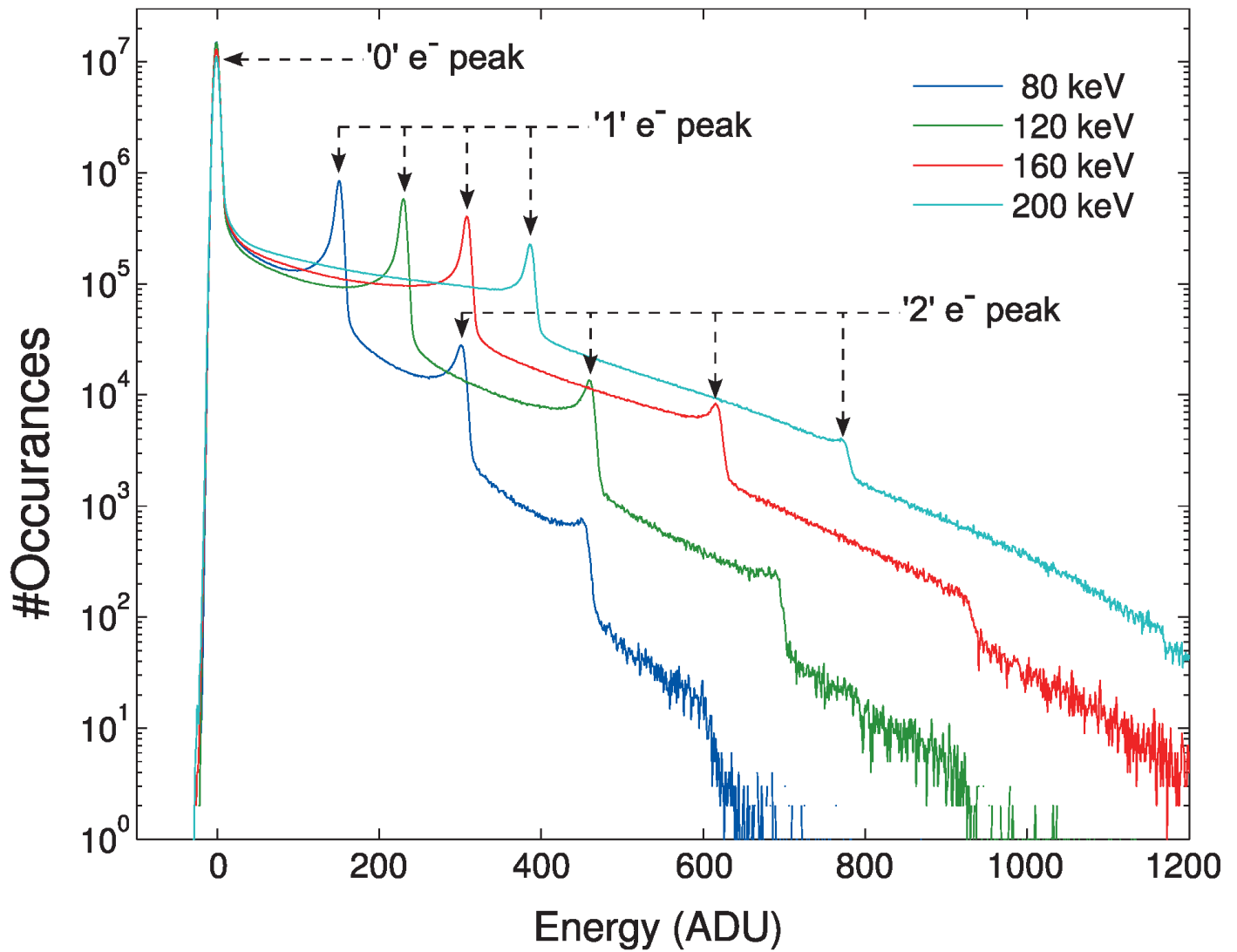

Figure 5. Pixel response histogram from 10,000 frames for very low, uniform illumination of the detector with electrons of 80, 120, 160, and 200 keV energy. The peak at 0 ADU corresponds to pixels with no electrons. For the 80 keV curve, the peaks at 151, 303, and 454 ADU correspond to peaks from 1, 2 and 3 microscope electrons per pixel. Intensity between the peaks arises from the signal of some electrons being shared between adjacent pixels. Discrete electron peaks for higher energy electrons occur at proportionately higher ADU values.

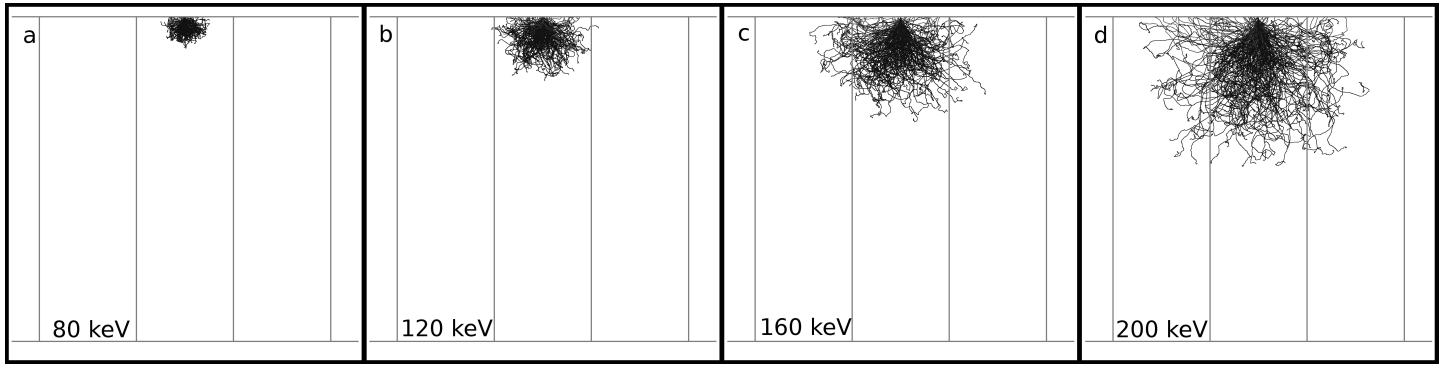

Figure 6. Monte Carlo simulations of electron tracks in silicon for a) 80, b) 120, c) 160, and d) 200 keV electrons. Each panel shows 200 individual electron tracks which impinge on the sensor at the top of each panel (3 pixel area covered in each panel). Pixel dimensions of 500 μm thickness (top to bottom) and 150 μm laterally are indicated by gray lines.

Fig 7

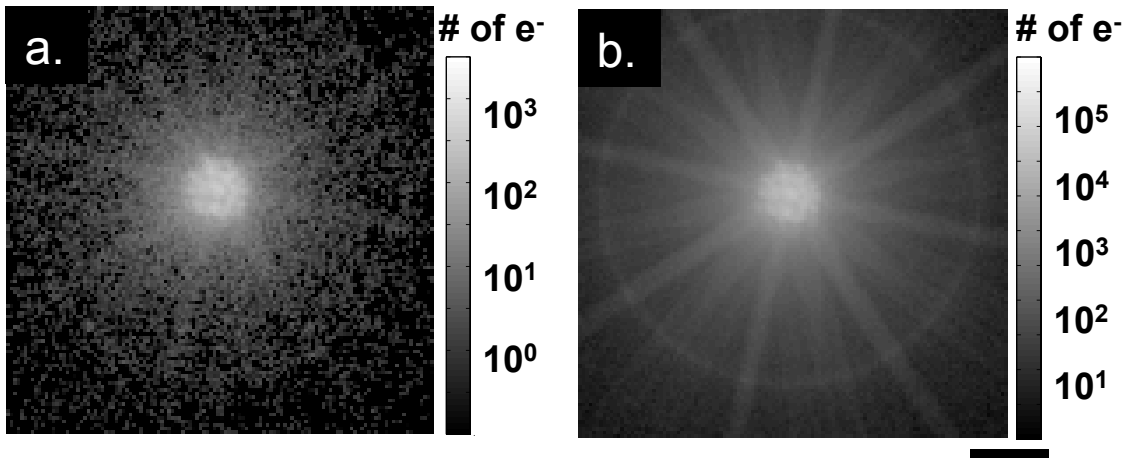

Fig 7: a.) CBED pattern of BiFeO$_3$ recorded in a) 1 ms and b) 100ms with 10 pA of beam current at a single scan position. All CBED images shows the number of primary electrons detected on the EMPAD detector, showing quantitative measurements of electron counting. Black bar on the lowermost far right represents 20 mrad for the diffraction patterns above.

Fig 8

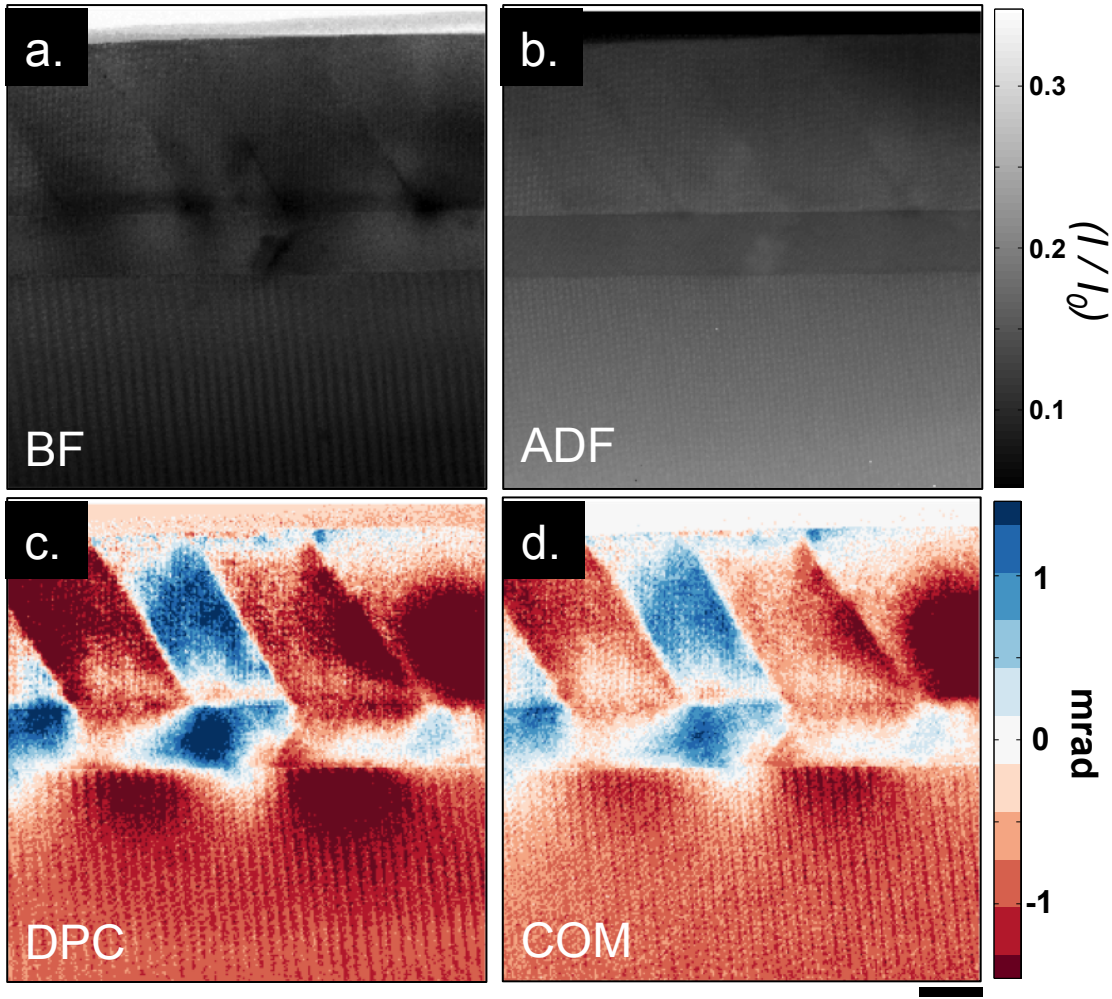

Fig 8. Different imaging modes extracted from a EMPAD-STEM image of 137 nm thick BiFeO$_3$ film grown on a 54 nm SrRuO$_3$ electrode on a DyScO$_3$ substrate. The acquisition time was 65 seconds. a.) Bright field signal from 0 to 5 mrad, b) annular dark field signal from 50-250 mrad, c) differential phase contrast in the x-direction and d) center of mass shift in the x-direction. COM deflections are shown in milliradians. The black scale bar under d), which is common to all the images is 50 nm. The (I/I$_0$) colorbar is common to the BF and HAADF images of panels a and b where is the I$_0$ incident beam intensity. The striations are aliasing artifacts from undersampling of the atomic lattice with an atomic-sized probe.

Fig 9

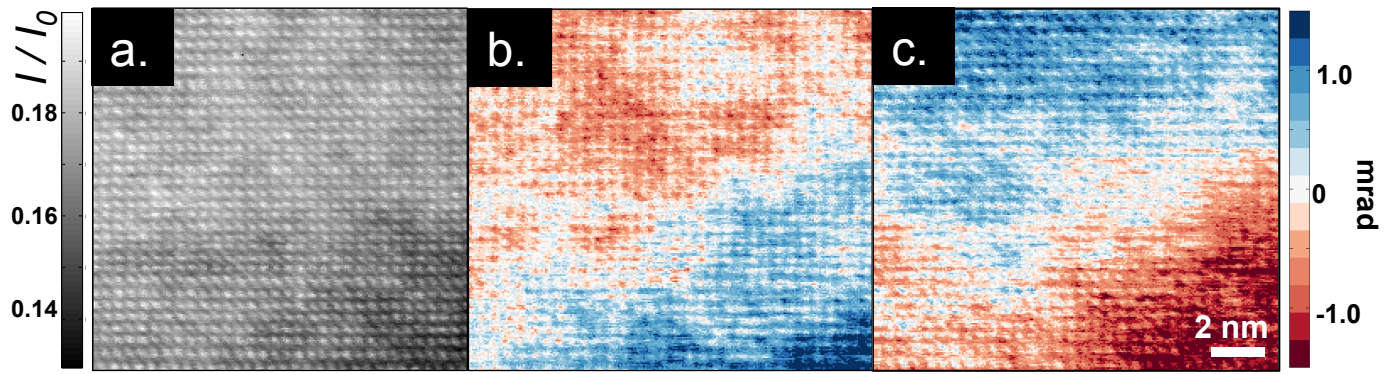

Fig 9: a) Atomic resolution HAADF image of BiFeO$_3$ taken using the EMPAD at a domain boundary. b) center of mass image taken in x and c) y showing structure changes at the atomic domain boundaries where deflections are shown in milliradians. The 2 nm scale bar in c also corresponds to a and b. All images are extracted from the same EMPAD-STEM data set, with 65 second acquisition time.

Fig 10

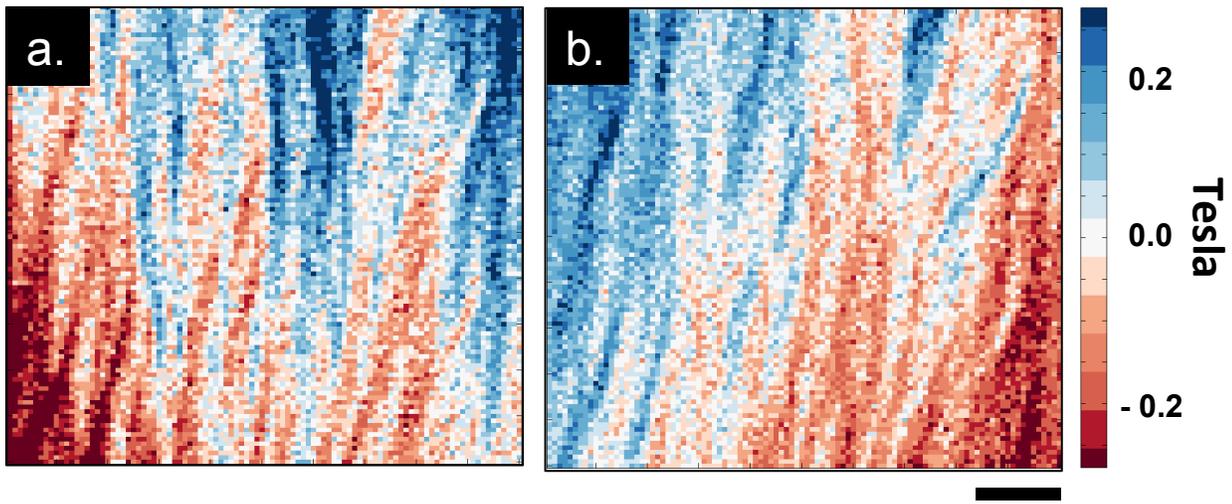

Fig 10: Center of Mass measurements of magnetic deflections (converted from mrad to Tesla) for 50 nm Co specimen on silicon nitride in a) x- and b) y-directions. Black scale bar represents 4 microns. Recorded in LM-STEM mode with the objective lens turned off for field free-imaging.

Fig 11

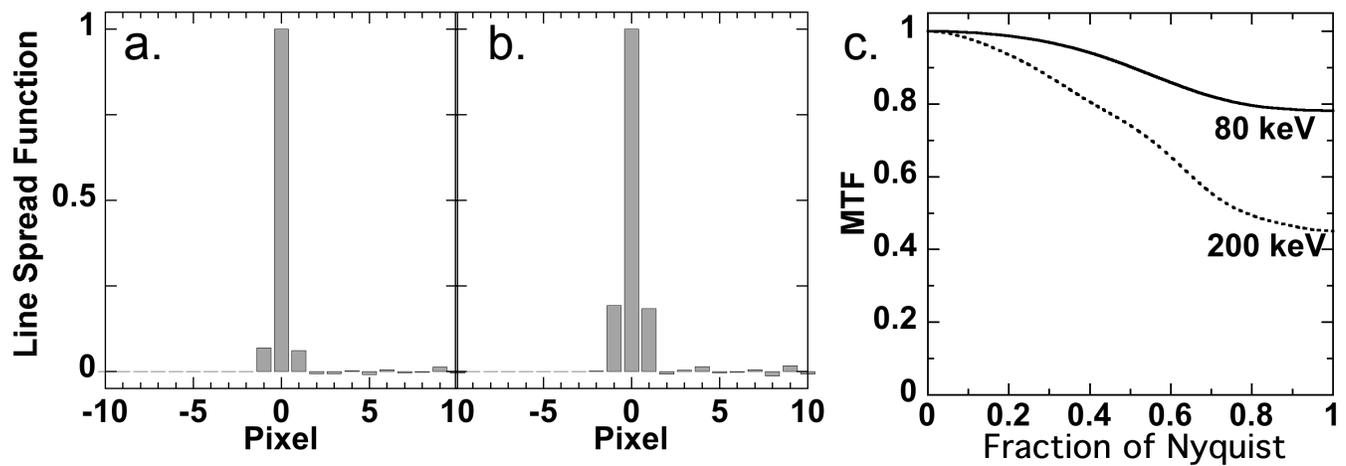

Fig 11: Measured detector response from the edge of an aperture imaged onto the detector. Line spread functions for (a) 80 and (b) 200 keV were calculated from the derivative of 100 images of the edge. The modulation transfer function (c) obtained from the Fourier transforms of the line spread functions, with spatial frequency plotted as a fraction of the Nyquist frequency.